%% file: fcc.tex
\documentclass[prl,aps,preprintnumbers,footnoteinbib,twocolumn]{revtex4-1}

\usepackage[T1]{fontenc}
\usepackage[]{natbib}
\usepackage{amscd}
\usepackage{amsmath, bbm} 
\usepackage{amsfonts}
\usepackage{amssymb}
\usepackage{slashed}
\usepackage{multirow}
\usepackage{array}
\usepackage{xcolor}
\usepackage{hyperref}
\usepackage[]{tabularx}

\usepackage{tikz}
\usepackage{tikz-feynman}
\tikzfeynmanset{compat=1.1.0}

\usepackage[]{siunitx}

\let\originalleft\left
\let\originalright\right
\renewcommand{\left}{\mathopen{}\mathclose\bgroup\originalleft}
\renewcommand{\right}{\aftergroup\egroup\originalright}

\newcommand{\bsll}{\ensuremath{b \rightarrow s \ell^+ \ell^-}}
\newcommand{\hc}{^{\dagger}} 
\newcommand{\flavio}{\texttt{flavio}}
\newcommand{\smelli}{\texttt{smelli}}

\newcommand{\pkg}[1]{\texttt{#1}}
\newcommand{\ee}{\ensuremath{e^+ e^-}}

\begin{document} 
\title{Computation of FCC-ee Sensitivity to Heavy New Physics with Interactions of Any Flavor Structure}
\author{Ben Allanach} \email{ben.allanach.work@gmail.com} \affiliation{DAMTP, University of Cambridge, Wilberforce Road, Cambridge, 
CB3 0WA, United Kingdom}
\author{Eetu Loisa} \email{eal47@cam.ac.uk} \affiliation{DAMTP, University of Cambridge, Wilberforce Road, Cambridge, 
CB3 0WA, United Kingdom}
\begin{abstract}
  We present a tool to compute the sensitivity of the
  Future Circular Electron--Positron Collider (FCC-ee) to the interactions of new, heavy
  particles 
   via publicly available extensions to the
   {\tt smelli} and {\tt flavio} computer programs. 
   We parameterize new particles' effects without any flavor
   assumptions and 
   take into account the projected experimental and correlated
   theoretical uncertainties of various electroweak and Higgs observables at
   the proposed collider.  
   We illustrate a use of the tool
   by estimating the sensitivity of the FCC-ee
   to a $Z^\prime$ model with flavor-specific couplings which explains 
   anomalies inferred from present-day measurements and Standard Model
   predictions of 
   observables that involve the $ \bsll $ transition.
\end{abstract}
\maketitle

\input{intro.tex}

\input{observables.tex}

\input{TFHM.tex}

\input{conclusions.tex}

\section{Acknowledgments}

We thank other members of the Cambridge Pheno Working Group for
discussions.
EL would like to thank Peter Stangl, Alex Mitov and Daniel Yeo for helpful suggestions. 
This work has been partially supported by STFC consolidated grants ST/X000664/1 and ST/T000694/1. 
EL is also supported by the Alfred Kordelin Foundation.

\appendix
\input{appendix.tex}


\bibliography{fcc}

\end{document}

%% file: intro.tex
The Future Circular Electron--Positron Collider (FCC-ee) is a proposed $e^+e^-$ collider envisaged to be based at CERN~\cite{FCC:2018evy} and
start collisions in the 2040s. 
It is designed to provide detailed studies of the four most massive particles of the Standard
Model (SM): $W^\pm$ bosons, Higgs bosons, $Z$ bosons 
and top quarks. 
The
FCC-ee affords a significant increase in the precision of measurements of
the properties of these heavy particles and sensitivity to rarer
decay modes. 
This enhanced precision enables the exploration of 
the effects of new particles that may have hitherto evaded detection due to
their high mass scales or small interaction strengths. 
Currently there is much activity to investigate the scientific benefits of the
FCC-ee in order to further motivate funding and building it.

We provide here a computational tool utilizing existing
estimates of the collider's experimental precision to aid this activity.
The tool estimates the sensitivity of the FCC-ee to extensions of the SM, making no assumptions about the flavor structure of new physics.
As such, it can be used to study highly flavorful new physics scenarios, and it 
can later be augmented with observables
deriving from the FCC-ee flavor physics program, on which there is ongoing research.

We will illustrate the use of our tool by quantifying the substantial
testing power the FCC-ee would provide on a flavorful new physics model which has been proposed to
explain aspects of the fermion mass problem and certain discrepancies between
measurements of $B$ meson decays and their SM predictions. We shall find that
the set of FCC-ee observables we 
include has the sensitivity to easily rule out such an explanation.  
This provides a concrete illustration of the potential power of the
FCC-ee in a flavored new physics context and adds to the FCC-ee physics
case. 

Our assumptions are that only one linearly realized Higgs field contributes significantly to electroweak symmetry breaking, that the masses of beyond-the-SM fields are
significantly greater than 365~GeV and that in the infinite mass limit 
all of those fields decouple.
Under such circumstances, the effects of new physics can be
characterized by the SM Effective Field Theory (SMEFT).
Using the language of the SMEFT allows us to leverage the \smelli~\cite{Aebischer:2018iyb} and \flavio~\cite{Straub:2018kue}
computer packages, which already contain the predictions and experimental
measurements of hundreds of $B$ meson, electroweak and other observables using
the SMEFT framework \footnote{\smelli\ accounts for the renormalization group running of the SMEFT operators by numerically integrating the $d=6$ renormalization 
group equations via the \texttt{wilson}
package~\protect\cite{Aebischer:2018bkb}.}. 
We shall extend these programs to include the estimated FCC-ee uncertainties of various Higgs and electroweak observables for which well-studied and official sensitivity estimates are available. 

\section{Standard Model Effective Field Theory}

The SMEFT encodes the effects of new physics in Wilson coefficients (WCs) -- dimensionless numbers multiplying local, irrelevant operators composed of the SM degrees of freedom and invariant under the SM gauge group.
The SMEFT Lagrangian density is 
\begin{equation}
{\mathcal L} = {\mathcal L}_4 + \sum_{d=5}^\infty \sum_i \frac{C_i}{\Lambda^{d-4}}
{\mathcal O}_i^{(d)}, 
\label{eq:Lsmeft}
\end{equation}
where ${\mathcal L}_4$ is the usual renormalizable SM Lagrangian.
The parameter $\Lambda$ stands for the SMEFT cut-off scale, usually taken to be the mass scale of heavy states that have been integrated out of an underlying renormalizable
field theory. 
The $C_i$ are the dimensionless WCs.
The index $i$ labels independent operators whereas $d$ is the canonical mass dimension of the operator ${\mathcal O}_i^{(d)}$. 

The only $d=5$ operator in the SMEFT expansion, the Weinberg operator \cite{Weinberg:1979sa}, describes neutrino masses and mixing.
At the $d=6$ level, the number of independent gauge invariant operators is large: 2499; these terms describe the physical effects that we are most interested in.
A judicious choice of a non-redundant operator basis is important to make sense of the expansion, and we shall henceforth adopt the operators and conventions of the Warsaw basis \cite{Grzadkowski:2010es}.
Effects of yet higher dimension are
suppressed by growing powers of $(E/\Lambda) \ll 1$, where $E$ is the energy scale of the physical process of interest. 
Since such higher $d$ operators are generically expected to make smaller impacts on 
observables, they are not accounted for in our approximation. 

In a specific SM extension with heavy new fields, many effective
operators are typically induced. 
The WCs of these operators are highly correlated with each other, being 
controlled by a small number of fundamental parameters.

%% file: observables.tex
\section{FCC-ee Observables}
This section aims to present and discuss the FCC-ee observables incorporated
in our newly developed extensions of \flavio{} \texttt{v2.6.1} and \smelli{} \texttt{v2.4.2}.
Where possible, we have followed the principles established in \cite{Falkowski:2019hvp}.
We adhere to the sensitivity estimates reported in the 2021 Snowmass proceedings \cite{Bernardi:2022hny, deBlas:2022ofj} and \cite{DeBlas:2019qco}.
The assumed running program corresponds to unpolarized electron and positron beams at center-of-mass energies $ \sqrt{s} = $ 91, 161, 240, 350 and 365 GeV with luminosities 150, 10, 5, 0.2 and 1.5 \unit{ab^{-1}}, respectively.
All projected measurements are assumed to be centered on the SM predictions.
 The FCC-ee observables are divided into two classes which we shall discuss in the following
sections: electroweak precision
  observables (EWPOs)
  and Higgs observables.



\subsection{Electroweak precision observables}
The FCC-ee would probe the electroweak sector at an unprecedented level of precision, improving upon the current measurements of many observables by two orders of magnitude.
Its planned runs at a variety of $\sqrt{s}$ values ensure experimental sensitivity to a host of SMEFT operators, ranging from four-fermion contact interactions to bosonic field strength operators.
Using the estimates of \cite{deBlas:2022ofj} for the FCC-ee measurement uncertainties,
we have included FCC-ee uncertainties for the key $ Z $-pole precision observables: the $ Z $ boson width, the total hadronic cross-section $ \sigma_\text{had}^0 $, the hadronic cross-section ratios $ R_f $ and left-right asymmetries $ A_f $.
Furthermore, the $ WW$, $ Zh $ and $ \overline{t} t $ runs allow for precise determinations of many $ W
$ boson observables, and we include FCC-ee uncertainties for the $ W $ boson mass $ M_W $, the $ W  $ boson width $ \Gamma_W $, as well as the inclusive $ WW $ production cross-sections and leptonic branching ratios $ \text{BR}\left( W \rightarrow \ell \nu \right) $ measured in the 161, 240 and 365 GeV runs of the collider.
The uncertainty estimates for the latter two sets of observables are taken from \cite{DeBlas:2019qco}.

The inclusive $ WW $ production cross-sections are simulated using the \pkg{MadGraph5\_aMC} \cite{Alwall:2014hca} event generator together with the \pkg{SMEFTsim} \cite{Brivio:2017btx,Brivio:2020onw} model files.
We use the built-in electroweak parton distribution functions \cite{Frixione:2021zdp} on \pkg{MadGraph} which account for initial state radiation (and beamstrahlung effects for $ \sqrt{s} = 240 $ and 365 GeV), turn on WCs one at a time and seek corrections to the SM cross-sections at linear order in the WCs.
This amounts to considering the interference terms between the SM amplitude and the SMEFT corrections
\begin{equation}
	\sigma = \sigma_{\text{SM}} + \sum_{i} a_i \frac{C_i}{\Lambda^2},
\end{equation}
where $ a_i $ gives the interference contribution to the cross-section.
We include in the cross-sections those SMEFT operators for which
\begin{equation}
	\frac{| a_i |}{\sigma_\text{SM}} > 10^3\, \text{GeV}^2.
\end{equation} 

Finally, we also include fermion scattering cross-sections and forward-backward asymmetries at $ \sqrt{s} = 240 $ and $ 365 $ GeV for the $ \ee $ $ \mu^+ \mu^- $, $ \tau^+ \tau^- $, $ \overline{c} c  $ and $ \overline{b} b $ final states, as reported in \cite{deBlas:2022ofj}.
These observables excel at probing four-fermion operators because their
interference with the SM grows with $\sqrt{s}$ \cite{Berthier:2015oma,Ge:2024pfn}, and their ability to test flavor non-universal models has recently been analyzed in \cite{Greljo:2024ytg}.
When applicable, we have cross-checked our analytic results with similar calculations in \cite{Greljo:2022jac,Allanach:2023uxz}.
The scattering amplitudes are calculated at tree-level including
$\mathcal{O}(\Lambda^{-2})$ corrections. 
We mod-square each amplitude to calculate the cross-section, thus including the effects of 
four-fermion operators that do not interfere with the SM amplitudes. 
The full set of electroweak observables is collected in Table~\ref{tab:ewpo_table}.

\subsection{Higgs measurements}

The two leading Higgs production modes at the FCC-ee are $ e^+e^- \rightarrow Zh $ (Higgstrahlung) and $ e^+e^- \rightarrow h \overline{\nu} \nu $ ($W$ boson fusion), where $h$ stands for the physical Higgs field. 
The third-most prevalent mode, $ Z $ boson fusion, is tenfold suppressed \cite{Bernardi:2022hny} relative to the first two at the FCC-ee energies \qty{240}{GeV} and \qty{365}{GeV} and is neglected.
The Higgs measurements are often reported in the form of signal strengths, $ \mu $,  defined as 
  \begin{equation}
	\mu \equiv \frac{[\sigma_i  \cdot \text{BR} \left( h \rightarrow f \right) ]_\text{observed} }{[\sigma_i  \cdot \text{BR} \left( h \rightarrow f \right)]_\text{SM}},
\end{equation}
for a given $h$ production mode cross-section $ \sigma_i$ and branching ratio $ \text{BR}\left( h \rightarrow f \right)  $ into final state $ f $.

\pkg{MadGraph} and \pkg{SMEFTsim} are employed to simulate the dominant Higgs production modes at the FCC-ee, following precisely the same procedure as outlined above for $ WW $ production.
These are then normalized to their SM predictions to derive signal strengths. 
The full set of Higgs observables can be found in Table~\ref{tab:higgs_table}.

\subsection{Treatment of theory uncertainties}
In order to translate the vast improvements in experimental precision at the
FCC-ee into tests of the SM, it is imperative that the theory errors of the
relevant observables are improved to match, or surpass, the experimental precision.
Both parametric theory errors, arising from the finite measurement precision of the SM input parameters, and intrinsic theory errors, emerging from missing higher order contributions in the SM predictions for various observables, must be controlled.
The feasibility of sufficiently large improvements before FCC-ee switch-on was the subject of \cite{Freitas:2019bre, Blondel:2019qlh}, where it was deemed that with a concerted effort the theory errors may be brought down to match the experimental precision for the observables considered in this letter.
Furthermore, explicit estimates for the parametric and theory uncertainties were listed.
We have incorporated these estimates into the FCC-ee observables introduced here. 
Where projected theoretical uncertainties for a given observable are unavailable, we assume the total theoretical uncertainty matches the projected experimental error, unless the current theoretical error is smaller, in which case we adopt the present-day value.

We add the parametric and intrinsic theory uncertainties to the projected
experimental uncertainties in quadrature, both for the EWPOs and
for the Higgs observables. This implicitly assumes that the three
sources of uncertainty are uncorrelated, which should hold to a reasonable
degree. Our procedure also implicitly treats the
theory uncertainty as a Gaussian random variable.
This makes our procedure operationally simple, as other treatments would
likely require dedicated changes to the core \flavio{} program.

All projected experimental uncertainties are treated as uncorrelated. 
As for the theory errors, whenever the predicted values of two observables are obviously correlated, we treat them as such.
Thus, for instance, we take into account that the cross-sections $
\sigma\left( \ee \rightarrow \mu^+ \mu^- \right)  $ and $ \sigma\left( \ee
\rightarrow \tau^+ \tau^- \right)  $ rely on the same theory prediction to a
good approximation, as do the Higgs signal strength predictions for a given production mode but different decay channels.
Where the correlation between two theory predictions is approximately unity, we treat the theory errors as fully correlated.
For another example, we therefore treat the theory errors in $ \sigma\left( \ee
  \rightarrow \mu^+ \mu^- \right) $ at $ \sqrt{s} = \qty{240}{GeV} $ and at $ \sqrt{s} = \qty{365}{GeV} $ as fully correlated.

The estimation and implementation of both projected experimental and theory errors is not an exact science.
We caution that our estimates are subject to refinement over the
  coming decades prior to FCC-ee operation.

%% file: TFHM.tex
\section{An illustration: sensitivity to the third family hypercharge model}
  We now demonstrate the new code with a use-case: that of estimating the
  sensitivity of the FCC-ee to a particular new physics model that explains
  discrepancies between certain measurements and SM predictions of
  $B$ meson
  decay observables. 
The Third Family Hypercharge Model (TFHM) \cite{Allanach:2018lvl} extends the
SM gauge group by a $ U(1)_{Y_3}$ factor under which the third family fermions
and the Higgs boson have charges proportional to their hypercharge, but under
which the other SM fields are uncharged.
$ U\left( 1 \right)_{Y_3} $ is spontaneously broken around the TeV scale by a SM singlet complex scalar field, the flavon, $ \theta $, which takes on a vacuum expectation value (VEV) $ v_\theta $. 
As a result, the $ Z^\prime $ gauge boson which mediates the new interaction acquires a mass, $ M_{Z^\prime} $.
Despite being coupled to only the \emph{third} generation of SM fermions in the
gauge eigenbasis, 
the $ Z^\prime $ acquires interactions with the lighter fermion species when
the fermions are rotated into the mass eigenbasis; an angle $\theta_{sb}$, for example, parameterizes the mixing between the left-handed strange and
bottom quark fields.
This allows the $Z^\prime$ field to mediate $ \bsll $ transitions via Feynman
diagrams that, after matching to the SMEFT, yield four-fermion operators
suppressed by the ratio $C_i / \Lambda^2 \sim g_{Z^\prime}^2 / M_{Z^\prime}^2 $
with $ g_{Z^\prime} $ being the $ U(1)_{Y_3}$ gauge coupling.
Furthermore, as the Higgs field $ H $ is charged under both $ U(1)_Y  $ and $ U(1)_{Y_3} $, the
$Z$ boson and the $Z^\prime$ mix with 
mixing angle $ \alpha_z $,
\begin{equation}
\sin \alpha_z = \frac{g_{Z^\prime}}{\sqrt{g_L^2 + g_Y^2} } \left( \frac{M_Z
}{M_{Z^\prime}} \right)^2 + {\mathcal O}\left(\frac{M_Z^4}{M_{Z^\prime}^4}\right),
\end{equation}
at tree-level.

To leading order, two parameters of the TFHM determine its ability to fit $ B $ meson decay data: the ratio $ g_{Z'} / M_{Z'} $ and $ \theta_{sb} $.
Whilst the $ \bsll $ anomaly landscape has evolved over the recent years (see \cite{Capdevila:2023yhq}), the ability of the TFHM to improve the fit to the anomalous $B$ meson decay measurements endures.
We use three sets of observables defined in \smelli\ to calculate
a global likelihood consisting of hundreds of current measurements.
The `Quarks' data set contains various rare $ B $ meson decay observables, some of which are in tension with the SM, as well as neutral meson mixing and other commonly studied flavor observables.
The data set `LFU FCNCs' consists of measurements testing lepton flavor universality, including the formerly anomalous $ R_K $ and $ R_{K^*} $, and `EWPOs' is made up of $ Z $- and $ W $-pole electroweak observables.
The best-fit point of the resulting fit improves upon the SM by 29.1 units of $\chi^2$~\footnote{For
$M_{Z^\prime}=3$ TeV, we find a best-fit point of 
$\theta_{sb} = -0.182$ and $g_{Z'} = 0.412$ for a normalization in which the
$U(1)_{Y_3}$ charge of the third-family left-handed quark doublet is $1/6$.
The results are summarized in Table~\ref{tab:fit_results}, which updates the fit presented in \cite{Allanach:2021kzj}.}.

\begin{table}[htpb]
	\centering
	\caption{
		The goodness of fit of the TFHM at its best-fit point, $ \{ g_{Z'} = 0.412, \, \theta_{sb} = -0.182 \}$, when $ M_{Z'} $ is set to \qty{3}{TeV}.
		From left to right, the columns show the names of the
                observable set used, the $ \chi^2 $ values, the number of
                observables in each set, the $p-$values and finally the improvements in $ \chi^2 $ relative to the SM fit, with positive values signalling an improved fit.
	}
	\label{tab:fit_results}
	\def\arraystretch{1.2}
	\begin{tabular}{ccccc}
	\hline\hline
	Data set & $ \chi^2 $ &$n$ & $ p $-value & $ \Delta \chi^2 $ \\
	\hline
	Quarks & 393.8 & 306 & $5.14 \times 10^{-4}$ & 29.45\\
	LFU FCNCs &19.0 &24 & 0.75 & $-0.43$ \\
	EWPOs & 36.9 & 31& 0.22 & 0.08\\
	\hline
	Global &449.4& 361 &$ 1.0 \times 10^{-3} $ & 29.10 \\
	\hline\hline
	\end{tabular}
\end{table}

Another point of interest is the scalar sector of the model. 
Aside from the mass terms and quartic self-interaction terms that generically exist for both the Higgs field $ H $ and the flavon field $ \theta $ in the Lagrangian, the symmetries of the model allow for a marginal interaction term $ \lambda_{H\theta} \left( H\hc H \right)\left( \theta^* \theta \right) $.
As a result, integrating out the heavy flavon from the theory yields a
contribution to the $d=6$ bosonic operator $ \left( H\hc H \right) \Box \left( H\hc H \right) $, which in the broken electroweak phase and under canonical normalization of the physical, real Higgs field results in the rescaling of all Higgs couplings in the SM (see for instance \cite{Alonso:2013hga}). 
This operator can also be viewed as capturing the mixing between the Higgs and
flavon fields, necessitating a rotation into a new scalar mass basis.
At tree-level, the mixing angle $\phi$ required to diagonalize the mass matrix obeys
\begin{equation} \label{eq:sin2phi}
	\sin 2 \phi = \frac{2 \lambda_{H \theta} v_H v_{\theta}}{m_{\theta}^2 - m_{h}^2},
\end{equation}
where $ v_\theta $ and $ v_H $ are the VEVs of the flavon and Higgs fields,
respectively, with $ m_\theta $ and $ m_h $ being their respective physical masses.


The $Z^\prime$ leaves an imprint on the EWPOs (a detailed discussion of the
fit to present-day EWPO data can be found in \cite{Allanach:2021kzj}), not
least because the custodial symmetry of the SM is now violated.
Furthermore, $ \phi $ influences the electroweak sector through one-loop effects, in addition to its tree-level impact on Higgs production processes. 
These features make the model susceptible to constraints from both electroweak
and Higgs production observables, measurable at current colliders as well as
at the FCC-ee.

\begin{figure}[htpb]
	\centering
	\includegraphics[width=8.5cm]{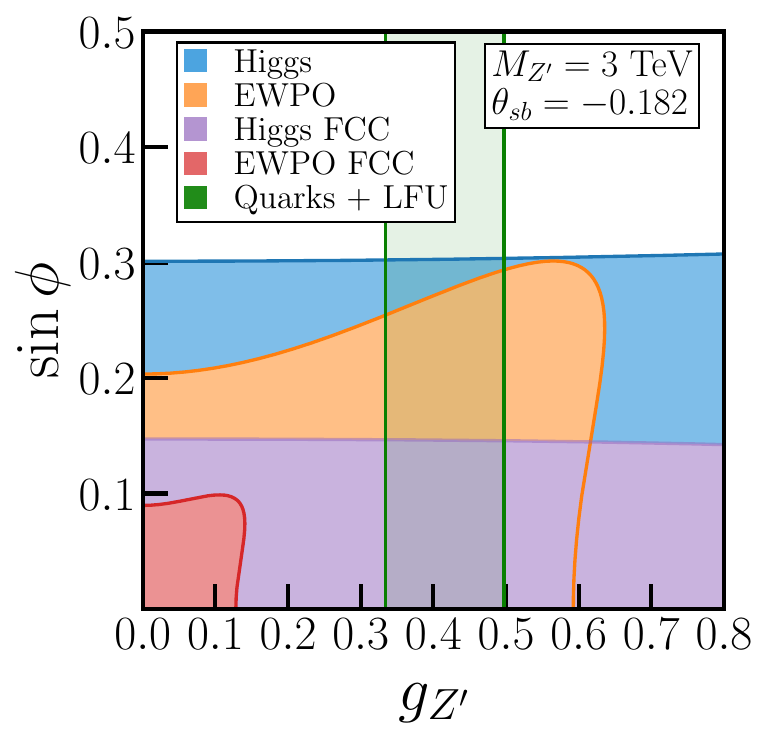}
	\caption{95\% confidence level (CL) contours of the TFHM in the
            $\sin \phi-g_{Z'}$ plane for $M_{Z'}=3$ TeV and $\theta_{sb}=-0.182$.
	    The plot shows current constraints and SM-centered FCC-ee projections.
          The colored region corresponding to the legend is `allowed' for
          each class of observables: that
          of `Quarks + LFU' shows the current 95$\%$ region preferred by the fit to
          $B$ meson data; `Higgs' and `EWPO' refer to current
          constraints. Legend items including the word `FCC' show FCC-ee sensitivity
          estimates. 
 }
	\label{fig:tfhm_plot}
\end{figure}

  Figure~\ref{fig:tfhm_plot} shows
  the expected improvement in precision of constraints on the model from Higgs
  observables and EWPOs at the FCC-ee in a plane of TFHM parameter space.
  The SM point lies at the origin: we see initially that the `Quarks + LFU' region,
  preferred by $B$ meson decay data,
  is far from this point, but that current Higgs and EWPO measurements are
  compatible with it. The figure also illustrates the improvement in precision
  and sensitivity once FCC-ee measurements are taken into account. Supposing
  the SM to be the theory chosen by Nature, the TFHM region of parameter space
  currently preferred by the `Quarks + LFU' data (which includes the anomalous $ B $ decays) would be strongly disfavored by the FCC-ee.

Both current and projected electroweak measurements show a mild degeneracy, where increasing $ g_{Z'} $ and $ \phi $ simultaneously allows for a mixing angle larger than when the gauge coupling is zero.
This is because an increased Higgs--flavon mixing angle acts to lower $ M_W $, whilst an increased $ g_{Z'} $ makes the $ W $ boson heavier.
It is these opposite sign contributions to $ M_W $, whose experimental world
average (prior to the 2022 CDF measurement) is in $ 2\sigma $ tension with the
SM \cite{ParticleDataGroup:2024cfk}, that give rise to this behavior.

%% file: conclusions.tex
\section{Closing remarks}

We have extended the \smelli{} and \flavio{} computer packages to provide an estimate of
  the FCC-ee sensitivity to heavy new physics with family-dependent couplings.
  We have accounted for the leading correlations among theoretical
  uncertainties which, to the best of our knowledge, is not implemented in other tools. 
  To illustrate the use of the new features, 
we have shown how the current TFHM fit can be strongly disfavored by electroweak and Higgs measurements at the FCC-ee. 
The FCC-ee is expected to deliver $10^{12}$ $B$ mesons~\cite{FCC:2018byv}, allowing for a further increase in precision on the measurement of processes involving the $b \rightarrow s \mu^+\mu^-$ transition. 
This may lead to the abandonment or further tweaking of the TFHM, since such processes were a large part of the initial motivation for it; in either case, indirect measurements at the FCC-ee will lead our direction beyond (or indeed back to) the SM. 

We hope to improve our computations in many ways in the future.
Several more observables could be added, for instance
\emph{differential} Higgs and $ W $ production observables, observables derived from $ t \overline{t} $ production and projected FCC-ee $B$ meson measurements, which are important in flavorful SM extensions.
It will also be of interest to study the correlations between various intrinsic
theory errors in greater detail and to include estimates of correlations among
the experimental uncertainties, for example from the luminosity measurement.
This uncertainty would have almost complete correlation across
all cross-section measurements and will be taken into account in future
versions of the program.
We note that a potential source of systematic uncertainty is from new
physics affecting the luminosity measurement
(which has a goal fractional precision of $10^{-4}$),
which then will filter into
other observables. At the FCC-ee, the luminosity will be inferred from small-angle
Bhabha scattering, $e^+e^- \rightarrow e^+e^-$~\cite{Dam:2021sdj}, which would be
affected by new physics that gives non-zero contributions to it. This will be
mitigated in future versions of the program by including the new physics
contribution to     low-angle Bhabha scattering.


The features introduced here could be straightforwardly extended to other proposed $ e^+ e^- $ colliders.
In addition to $ \ee $ colliders comparable to the FCC-ee, a similar analysis
could be conducted for the proposed TeV-scale muon colliders, although the
domain of validity of the SMEFT approach would be smaller because a muon collider
would operate at a larger center-of-mass energy, which the mass of new degrees of
freedom must be larger than. 
We believe that high-energy physics stands to benefit greatly from the development of accessible and automated tools for comparing the sensitivities of various proposed colliders to a range of concrete extensions of the SM.
In this letter, we hope to have taken a meaningful step towards that goal.
The computer programs and instructions for their use can be found at
\url{https://github.com/eetuloisa/smelli_fcc} and \url{https://github.com/eetuloisa/flavio_fcc}.

%% file: appendix.tex

\section{List of FCC-ee observables}
Tables~\ref{tab:ewpo_table} and \ref{tab:higgs_table} of this appendix display
the full sets of Higgs and electroweak observables included in our tool. 
Whenever an observable is defined at many different center-of-mass energies, $ \sqrt{s}  $ is treated as an extra parameter for the EWPOs, whereas $ \sqrt{s} $ appears in the name of the Higgs signal strength observables.
This feature, which results from striving to modify the existing \flavio{} program as minimally as possible, explains the differences in the naming conventions between the two tables.

\begin{widetext}
\begin{table*}[htpb]
	\centering
	\caption{The electroweak precision observables in \pkg{flavio} and \pkg{smelli} whose projected measurements have been added to the programs.
	The first column shows the name of the observable, the second its name in the code and the third a brief description.
}
	\label{tab:ewpo_table}
	\renewcommand{\arraystretch}{1.4}
	\begin{tabularx}{\textwidth}{cc@{\hskip .3cm}c@{\hskip .3cm}X}
		\hline
		\hline
		&Observable & \texttt{Name in program} & Description  \\
		\hline
	\multirow{12}{*}{\rotatebox{90}{$ Z $-pole observables}}	&$ \Gamma_Z $ & \texttt{GammaZ} & Total $ Z $ boson decay width \\
		&$ \sigma^0_\text{had} $ & \texttt{sigma\_had} & Cross-section $ \sigma (\ee \rightarrow \text{hadrons} ) $ at the $ Z $-pole  \\
		&$ A_e $ & \texttt{A(Z->ee)} & Left-right asymmetry in $ Z \rightarrow e^+ e^- $ decays \\
		&$ A_\mu $ & \texttt{A(Z->mumu)} & Left-right asymmetry in $ Z \rightarrow \mu^+ \mu^- $ decays\\
		&$ A_\tau $ & \texttt{A(Z->tautau)} & Left-right asymmetry in $ Z \rightarrow \tau^+ \tau^- $ decays\\
		&$ A_b $ & \texttt{A(Z->bb)} & Left-right asymmetry in $ Z \rightarrow \overline{b} b $ decays\\
		&$ A_c $ & \texttt{A(Z->cc)} &Left-right asymmetry in $ Z \rightarrow \overline{c} c $ decays \\
		&$R_e$ & \texttt{R\_e} & Partial decay width $ \Gamma_{Z\rightarrow e e} $ relative to hadronic width  \\
		&$R_\mu$ & \texttt{R\_mu} & Partial decay width $ \Gamma_{Z\rightarrow \mu \mu} $ relative to hadronic width\\
		&$R_\tau$ & \texttt{R\_tau} & Partial decay width $ \Gamma_{Z\rightarrow \tau \tau} $ relative to hadronic width\\
		&$ R_b $ & \texttt{R\_b} & Partial decay width $ \Gamma_{Z\rightarrow b b} $ relative to hadronic width\\
		&$ R_c $ & \texttt{R\_c} & Partial decay width $ \Gamma_{Z\rightarrow c c} $ relative to hadronic width\\
		\hline
		\multirow{10}{*}{\rotatebox{90}{Super-$ Z $-pole fermion scattering}}&$ \sigma\left( \ee \rightarrow \ee \right)  $ & \texttt{sigma(ee->ee)(high\_E)} & Cross-section of $ \ee \rightarrow \ee $; $ \sqrt{s} = 240, 365 $ GeV \\
		&$ A_\text{FB} (\ee \rightarrow \ee) $ & \texttt{AFB(ee->ee)(high\_E)} & Forward-backward asymmetry in $ e^+ e^- \rightarrow e^+ e^- $;  $ \sqrt{s} = 240, 365 $ GeV \\
		&$ \sigma\left( \ee \rightarrow \mu^+ \mu^- \right)  $ & \texttt{sigma(ee->mumu)(high\_E)} & Cross-section of $\ee \rightarrow \mu^+ \mu^- $; $ \sqrt{s} = 240, 365 $ GeV \\
		&$ A_\text{FB} (\ee \rightarrow \mu^+ \mu^-) $ & \texttt{AFB(ee->mumu)(high\_E)} &Forward-backward asymmetry in $ e^+ e^- \rightarrow \mu^+ \mu^- $; $ \sqrt{s} = 240, 365 $ GeV  \\
		&$ \sigma\left( \ee \rightarrow \tau^+ \tau^- \right)  $ & \texttt{sigma(ee->tautau)(high\_E)} &Cross-section of  $\ee \rightarrow \tau^+ \tau^- $; $ \sqrt{s} = 240, 365 $ GeV \\
		&$ A_\text{FB} (\ee \rightarrow \tau^+ \tau^-) $ & \texttt{AFB(ee->tautau)(high\_E)} &Forward-backward asymmetry in $ e^+ e^- \rightarrow \tau^+ \tau^- $; $ \sqrt{s} = 240, 365 $ GeV  \\
		&$ \sigma\left( \ee \rightarrow \overline{b} b \right)  $ & \texttt{sigma(ee->bb)(high\_E)} & Cross-section of $\ee \rightarrow \overline{b} b $; $ \sqrt{s} = 240, 365 $ GeV \\
		&$ A_\text{FB} (\ee \rightarrow  \overline{b} b) $ & \texttt{AFB(ee->bb)(high\_E)} & Forward-backward asymmetry in $ e^+ e^- \rightarrow \overline{b} b $; $ \sqrt{s} = 240, 365 $ GeV   \\
		&$ \sigma\left( \ee \rightarrow  \overline{c} c \right)  $ & \texttt{sigma(ee->cc)(high\_E)} & Cross-section of $\ee \rightarrow \overline{c} c $; $ \sqrt{s} = 240, 365 $ GeV \\
		&$ A_\text{FB} (\ee \rightarrow  \overline{c} c) $ & \texttt{AFB(ee->cc)(high\_E)} &  Forward-backward asymmetry in $ e^+ e^- \rightarrow \overline{c} c $; $ \sqrt{s} = 240, 365 $ GeV  \\
		\hline
		\multirow{7}{*}{\rotatebox{90}{$ W $ boson observables}}&$ M_W$ & \texttt{m\_W} & Pole mass of the $ W $ boson  \\
		&$ \Gamma_W $ & \texttt{GammaW} & Width of the $ W $ boson \\
		&$ R \left( \ee \rightarrow W^+ W^- \right)  $ & \texttt{R(ee->WW)} & Inclusive $ WW $ production cross-section normalized to SM prediction; $ \sqrt{s}~=~161,~240,~365$~GeV \\
		&$ \text{BR}\left( W \rightarrow e \nu \right)  $ & \texttt{BR(W->enu)}& $ W $ boson branching ratio into $ e \nu $ \\
		&$ \text{BR}\left( W \rightarrow \mu \nu \right)  $ & \texttt{BR(W->munu)}& $ W $ boson branching ratio into $ \mu \nu $\\
		&$ \text{BR}\left( W \rightarrow \tau \nu \right)  $ & \texttt{BR(W->taunu)}&$ W $ boson branching ratio into $ \tau \nu $ \\
			\hline \hline
	\end{tabularx}
\end{table*}
\begin{table*}[htpb]
	\centering
	\caption{The Higgs signal strength observables whose projected measurements were added into the programs.
	The first column shows the name of the observable, whereas the second column gives its name in the code. 
	The notation $ \{240 | 365 \} $ means that the observable is defined for two values of $ \sqrt{s}  $, one of which should be specified by the user.
}
	\label{tab:higgs_table}
	\renewcommand{\arraystretch}{1.4}
	\setlength{\tabcolsep}{12pt}
	\begin{tabular}{ll}
		\hline
		\hline
		Observable & \texttt{Name in program}  \\
		\hline 
		$ \mu \left(\ee \rightarrow Zh\right)  $& \texttt{mu\_Zh\_\{240|365\}(h->inc)}   \\
		$ \mu \left(\ee \rightarrow Zh; h \rightarrow \overline{b}b \right)  $& \texttt{mu\_Zh\_\{240|365\}(h->bb)}   \\
		$ \mu \left(\ee \rightarrow Zh; h \rightarrow \overline{c}c \right)  $& \texttt{mu\_Zh\_\{240|365\}(h->cc)}  \\
		$ \mu \left(\ee \rightarrow Zh; h \rightarrow gg \right)  $& \texttt{mu\_Zh\_\{240|365\}(h->gg)}  \\
		$ \mu \left(\ee \rightarrow Zh; h \rightarrow ZZ \right)  $& \texttt{mu\_Zh\_\{240|365\}(h->ZZ)}  \\
		$ \mu \left(\ee \rightarrow Zh; h \rightarrow WW \right)  $& \texttt{mu\_Zh\_\{240|365\}(h->WW)}  \\
		$ \mu \left(\ee \rightarrow Zh; h \rightarrow \tau \tau \right)  $& \texttt{mu\_Zh\_\{240|365\}(h->tautau)}  \\
		$ \mu \left(\ee \rightarrow Zh; h \rightarrow \gamma \gamma \right)  $& \texttt{mu\_Zh\_\{240|365\}(h->gammagamma)}  \\
		$ \mu \left(\ee \rightarrow Zh; h \rightarrow Z \gamma \right)  $& \texttt{mu\_Zh\_240(h->Zgamma)}  \\
		$ \mu \left(\ee \rightarrow Zh; h \rightarrow \mu \mu \right)  $& \texttt{mu\_Zh\_\{240|365\}(h->mumu)}  \\
		$ \mu \left(\ee \rightarrow h \nu \nu; h \rightarrow \overline{b} b \right)  $& \texttt{mu\_hnunu\_\{240|365\}(h->bb)}  \\
		$ \mu \left(\ee \rightarrow h \nu \nu; h \rightarrow \overline{c} c \right)  $& \texttt{mu\_hnunu\_365(h->cc)}  \\
		$ \mu \left(\ee \rightarrow h \nu \nu; h \rightarrow g g \right)  $& \texttt{mu\_hnunu\_365(h->gg)}  \\
		$ \mu \left(\ee \rightarrow h \nu \nu; h \rightarrow ZZ \right)  $& \texttt{mu\_hnunu\_365(h->ZZ)}  \\
		$ \mu \left(\ee \rightarrow h \nu \nu; h \rightarrow WW \right)  $& \texttt{mu\_hnunu\_365(h->WW)}  \\
		$ \mu \left(\ee \rightarrow h \nu \nu; h \rightarrow \tau \tau \right)  $& \texttt{mu\_hnunu\_365(h->tautau)}  \\
		$ \mu \left(\ee \rightarrow h \nu \nu; h \rightarrow \gamma \gamma \right)  $& \texttt{mu\_hnunu\_365(h->gammagamma)}  \\
		$ \mu \left(\ee \rightarrow h \nu \nu; h \rightarrow \mu \mu \right)  $& \texttt{mu\_hnunu\_365(h->mumu)}  \\
			\hline \hline
	\end{tabular}
\end{table*}
\end{widetext}